\documentclass{article}

\usepackage[T2A]{fontenc}
\usepackage[english,main=russian]{babel}
\usepackage{inputenc}

\usepackage{graphicx}

\usepackage{amsmath}
\usepackage{amssymb}
\usepackage{amsfonts}

\usepackage{csquotes}
\DeclareMathOperator{\divi}{div}

\newcommand{\hh}{h_{11}}
\newcommand{\hu}{H_{un}}

\newcommand{\apj}{ApJ}

\newcommand{\mnras}{MNRAS}

\newcommand{\prd}{Phys. Rev. D}

\author{A.\,N.\,Baushev \vspace{1cm}\\
 Bogoljubov Laboratory of Theoretical Physics, Joint Institute for Nuclear Research,
Dubna, Russia \thanks{baushev@gmail.com}\\
 Department of Physics, University of Helsinki, Helsinki, Finland}

\title{Spherization of the velocity field of cosmological perturbations near density extrema}

\date{}

\begin{document}
\setcounter{topnumber}{1}

\maketitle

\begin{abstract}
In this work, we consider a mechanism that tends to make the velocity field of linear cosmological perturbations near density extremum points (maxima or minima) spherically symmetric. It is shown that this mechanism begins to operate from the moment the Universe enters the stage of nonrelativistic matter domination ($z<z_{eq}\simeq 3600$) and rather quickly makes the velocity distribution near the density extremum practically isotropic. However, if the density profile of the initial perturbation is noticeably aspherical, its gravitational field subsequently destroys this isotropy.
\end{abstract}

{\textsc  PACS: 98.80.-k, 98.62.Js, 95.35.+d, 98.65.Dx}

\normalsize

\section{Introduction} 
According to modern concepts, all the structures that we observe in the present-day Universe evolved from initial cosmological perturbations, which constituted a random Gaussian field. Therefore, the density profile of these perturbations was quite complex and contained small-scale \enquote{roughness}. Ultimately, however, we are interested in observable astronomical objects (for example, galaxies) that have a certain characteristic mass. They form from perturbations that have a present-day length $\lambda_0$ corresponding to this mass $M\sim \frac43 \pi (\lambda_0/2)^3 \rho_{m0}$ (here and below, by the index \enquote{0}  we will denote the value of quantities in the present-day epoch). As long as the perturbations are linear, we can mentally get rid of the short-wavelength \enquote{ripple} that is unimportant for us. To do this, one needs to smooth the short-wavelength oscillations in both the density distribution and the velocity distribution; mathematically, this can be done by applying some high-frequency filter (for example, Gaussian smoothing). In what follows, without loss of generality, we will consider precisely such a smoothed profile: the evolution of perturbations of the masses of interest to us will proceed in the linear stage in exactly the same way as in the case of a real, unsmoothed profile.

A decisive role in the formation of present-day astronomical structures was played by dark matter~\cite{gorbrub2}. From the moment when it became nonrelativistic and practically ceased to interact with photons and baryonic matter (and this, obviously, happened in the early Universe) until the moment when the Universe exited the radiation-dominated stage, the density perturbations of dark matter that had entered under the horizon did not oscillate and grew (although very slowly, logarithmically). At the same time, the relative density perturbations of the baryon-photon component, being initially equal to the perturbations of dark matter, did not grow at all during the radiation-dominated stage, and by the moment of recombination the relative density contrast was tens of times smaller for baryons than for dark matter. Therefore (and also because dark matter is approximately 5 times greater in mass than baryons), it was precisely the distribution of dark matter perturbations that determined the present-day large-scale structure of the Universe: baryons, in essence, merely followed the gravitational potential created by dark matter. Therefore, in this work we will consider mainly precisely the perturbations of dark matter.

When considering structure formation, the greatest interest, as a rule, is aroused by the density maxima of the initial perturbations, since it is into them that matter flows during the growth of perturbations, and it is there that the brightest, and therefore easily observable, objects form. Our consideration and conclusions, however, will be equally applicable in the case of density minima as well. Thus, we will consider the neighborhoods of density extremum points.

Let us consider a region of the initial distribution from which the structure of interest to us will later develop (a density peak in the case of a galaxy or a galaxy cluster, or a density trough in the case of a void). Let us choose the point of the corresponding density extremum as the origin of coordinates and surround it with a sphere of comoving radius  $\tilde R\ll \tilde \lambda\equiv \lambda_0$. The index $\tilde .$ everywhere denotes a size comoving with the expansion of the Universe as a whole: a sphere of radius $\tilde R$ has radius $R$ at the present time and radius $R (z+1)^{-3}$ in the epoch with redshift $z$.  In the most general case, the distribution of density and velocity of matter inside the sphere $\tilde R$ up to terms of order $(\tilde R/\tilde \lambda)^3$ has the form 
\begin{equation}
\rho_M =\it{const}+\zeta_{ij} r_i r_j; \qquad v_i=\sum_j H_{ij} r_j.\label{27z1}
\end{equation}
Generally speaking, the density maxima and minima corresponding to a random Gaussian field have an elliptical shape with a characteristic axis ratio of $1:2-1:3$, but among them there is statistically also a certain fraction of nearly spherically symmetric density perturbations~\cite{bardeen1986}. In this work we will consider precisely such perturbations (we emphasize that we assume spherical symmetry of the density perturbation, but not of the velocity!), despite the fact that their fraction is small. The reason is that, although the velocity-field spherization mechanism presented below takes place for an extremum of arbitrary shape (albeit with different final consequences, see \emph{Discussion}), it is easiest to demonstrate its action precisely in the spherically symmetric case.

We emphasize that the spherical symmetry of the density distribution by no means implies spherical symmetry of the perturbation itself: the velocity distribution in such a perturbation is by no means obliged to be spherically symmetric. This is already evident from the circumstance that it is only the individual perturbation under consideration that possesses spherical symmetry of density, and not the distribution of matter in the entire Universe as a whole.

As an illustration, let us consider (deeply in the linear stage) a perturbation of characteristic size $\lambda$, in which not only the density distribution but also the velocity distribution is spherically symmetric. Let us introduce a coordinate system centered at the density extremum point. Let us add to the spherically symmetric perturbation two more adiabatic perturbations
\begin{eqnarray}
\Delta\rho =\mathrm{\rho} \cos(2\pi x/\Lambda)\qquad &\Delta v =-\mathrm{v} \sin(2\pi x/\Lambda)\nonumber\\
\Delta\rho=-\mathrm{\rho} \cos(2\pi y/\Lambda)\qquad &\Delta v =\mathrm{v} \sin(2\pi y/\Lambda),\label{27b1}
\end{eqnarray}
where $\Lambda\gg \lambda$. Let us consider the resulting perturbation. The addition of two plane waves will practically not change the density profile of the original perturbation: the two plane adiabatic waves practically (up to terms $\sim (\lambda/\Lambda)^2$) compensate each other. However, the velocity distribution in the resulting perturbation will no longer be spherically symmetric: the velocity components~(\ref{27b1}) lead to additional compression along the X axis and expansion along the Y axis. Thus, the velocity distribution even in a spherically symmetric density perturbation can be quite arbitrary. The presence of an asymmetry in the velocity distribution should lead to the gradual emergence of an asymmetry in the density distribution, but in the linear stage of perturbation growth this effect is small.

In this article we will show that there exists a mechanism that tends to spherize the velocity distribution near density extremum points (maxima or minima). In the case where the density profile of the halo is spherically symmetric, after the Universe transitions to the stage of nonrelativistic matter domination ($z<z_{eq}\simeq 3600$), the velocity field rapidly becomes fully isotropic.

\section{Calculations}
The case of a spherically symmetric density perturbation is convenient precisely because it is not difficult to show (see, for example, ~\cite{bondi1947,24,25}) that the evolution of the distribution of matter inside $\tilde R$ will, in the linear stage, proceed practically independently of the rest of the Universe. Indeed, the density near the extremum is practically (up to terms of order $(\tilde R/\tilde \lambda)^2$) constant, so there are no pressure gradients inside $\tilde R$ or at its boundary. In general relativity, as in Newtonian gravity, a spherically symmetric layer does not create a gravitational field inside itself (Birkhoff's theorem), and therefore the homogeneous Universe outside $\tilde R$ has no influence inside this sphere. The gravitational field there can be created only by asymmetric density perturbations outside $\tilde R$,  but all perturbations in the linear stage are small in general, and this effect is nonlinear in the perturbation amplitude and therefore negligibly small. As we have already noted, in the most general case the distribution of matter velocities inside the sphere $\tilde R$  is linear, i.e., has the form 
\begin{equation}
\label{27a0}
v_i=\sum_j H_{ij} r_j, 
\end{equation}
where $r_j$ is the radius vector drawn from the origin, and the quantities $H_{ij}$ depend only on time $t$. Therefore, from the continuity equation it follows that 
\begin{equation}
\label{27a1}
-\frac{\dot\rho}{\rho}=\divi \vec v=\sum_i H_{ii}
\end{equation}
also depends only on $t$, and the homogeneity of matter inside $\tilde R$ is preserved over time. We arrive at an important conclusion: just as the Tolman solution for a dust sphere behaves as a piece cut out of a Friedmann universe~\cite{tolman1934}, the cosmological solution inside $\tilde R$ behaves as a piece cut out of a homogeneous but anisotropic cosmological model. For brevity, let us call it a mini-universe and consider the evolution of its properties.

Let us rotate the reference frame so that the \enquote{Hubble tensor} $H_{ij}$ becomes diagonal. Let us introduce new notation:
\begin{equation}
\label{27a2}
H\equiv \frac{H_{11}+H_{22}+H_{33}}{3};\quad h_{ij}\equiv H_{ij}-H\delta_{ij}
\end{equation}
where $\delta_{ij}$ is the Kronecker symbol. The tensor $h_{ij}$ is diagonal, and $h_{11}+h_{22}+h_{33}=0$. Now the \enquote{Hubble tensor} can be represented as the sum of two components:
\begin{equation}
\label{27a3}
H_{ij}=H\delta_{ij}+h_{ij}
\end{equation}
The first component in~(\ref{27a3}) describes the isotropic component of the velocity field, which includes the sum of the expansion of the homogeneous Universe and the isotropic component of the perturbation. The component $h_{ij}$ describes the anisotropic part of the perturbation; since the perturbation is small, $h_{ij}\ll H$.

First let us consider the case when $h_{ij}=0$. Then our mini-universe is not only homogeneous but also isotropic, i.e., simply Friedmannian. Let us introduce a Friedmannian comoving reference frame (therefore the dark matter velocity in it is everywhere zero). In this reference frame the metric has the standard Friedmannian form $ds^2=dt^2-a^2(t) dl^2$, where $a(t)$ is the scale factor, and $H=\frac{1}{a}\frac{da}{dt}$. We will assume that at the present time $a_0=1$, then the scale factor $a$  is related to redshift by the simple relation $a=(z+1)^{-1}$. The evolution of $H$ and of the density is described by the usual Friedmann equations. 

What will change in this picture if the anisotropic component $h_{ij}\ne 0$? Since $h_{11}+h_{22}+h_{33}=0$, it follows from~(\ref{27a1}) that the evolution of density with time, and hence also the gravitational field, the accelerations of all particles, and the dependence $H(t)$ will not change at all compared with the isotropic case. We can use the same Friedmannian reference frame, isotropically expanding with constant Hubble parameter $H$, but in the case $h_{ij}\ne 0$ it will no longer be comoving: dark matter will move relative to it with velocities   
\begin{equation}
\label{27a4}
v_i=\sum_j h_{ij} r_j,
\end{equation}
determined by the anisotropic component of the perturbation. It is only this circumstance that distinguishes a universe with $h_{ij}\ne 0$ from a Friedmannian one with the same composition and the same constant $H$.

From this it is easy to find the law of evolution of the anisotropic part of the perturbation $h_{ij}(t)$. Since the accelerations of all particles coincide with the isotropic case, let us use the well-known property of a nonrelativistic particle moving in a Friedmannian universe: the product $\vec v\cdot a(t)=\it{const}$. Without loss of generality, we will assume that $\hh \ne 0$. Let us consider the point with comoving coordinates $[\tilde R, 0, 0]$,  i.e., the point of intersection of the X axis with the sphere $\tilde R$ (recall that the origin of coordinates is at the density extremum point). The velocity of dark matter at this point is directed radially, along the X axis, and is equal to $v=\hh(t) a(t) \tilde R$ at time $t$. At the next moment $t+dt$ the matter at this point will be that which at time $t$ was located at radius $a(t) \tilde R - v dt=a(t) \tilde R (1-\hh(t) dt)$, and therefore its velocity was equal to 
\begin{equation}
\label{27a5}
\tilde R a(t) \hh(t) [1-\hh(t) dt]
\end{equation}
(everywhere in this paragraph we neglect terms of second and higher order of smallness). This velocity has undergone a redshift, which can be found from the relation $v\cdot a(t)=\it{const}$: hence $a dv + v da=0$ and 
\begin{equation}
\label{27a6}
dv=-v\frac{da}{a dt} dt=-\tilde R a(t) \hh(t) H(t) dt.
\end{equation}
Combining~(\ref{27a5}) and~(\ref{27a6}), we obtain that at time $t+dt$ the velocity at the point $[\tilde R, 0, 0]$ is equal to 
\begin{equation}
\label{27a7}
v(t+dt)=\tilde R a(t) \hh(t) [1 - \hh(t) dt - H(t) dt].
\end{equation}
On the other hand,
\begin{eqnarray}
\label{27a8}
v(t+dt)=\tilde R a(t+dt) \hh(t+dt)=\\ \nonumber
=\tilde R a(t) \hh(t) [1 + \frac{d\hh(t)}{\hh(t) dt} dt + H(t) dt]
\end{eqnarray}
(here we have used that $a(t+dt)= a(t) (1+ H(t)dt)$). Equating~(\ref{27a7}) and~(\ref{27a8}), we obtain
\begin{equation}
\label{27a9}
\frac{d\hh}{\hh dt}= -2H -\hh.
\end{equation}
Since $\hh\ll H$, we can neglect the second term on the right-hand side. 
\begin{equation}
\label{27a10}
\frac{d\hh}{\hh dt}= -2H = -2\frac{da}{a dt}.
\end{equation}
From this,
\begin{equation}
\label{27a11}
\hh\propto a^{-2}(t).
\end{equation}
Clearly, $h_{22}$ and $h_{33}$ behave in exactly the same way. Thus, independently of the expansion regime of the Universe, the anisotropic component $h_{ij}(t)$ decreases as $a^{-2}(t)$. 

It remains to compare this result with the rate of decrease of the isotropic part of the perturbation $|H-\hu|$, where $\hu$ is the Hubble constant of the unperturbed Universe. Indeed, expressions~(\ref{27a0}) and~(\ref{27a3}) can be rewritten in the form
\begin{equation}
\label{27a12}
\vec v = \hu \vec r + (H-\hu) \vec r + \overrightarrow{h_{ij} r_j}
\end{equation}
Here the first term of the sum describes the unperturbed Universe (the velocity field specified by this term is always spherically symmetric), the second describes the spherically symmetric part of the perturbation, and the third describes the asymmetric part. 

Let us expand the perturbation into individual Fourier harmonics and take into account that the velocity perturbation is zero at the density extremum point, since the density and velocity perturbations are shifted in phase by $\pi/2$ in the case of adiabatic perturbations. Therefore each individual harmonic for the velocity has the form $v=v_a \sin(2\pi r/\lambda)$, where $v_a$ is its amplitude.

In the radiation-dominated stage $\hu \propto a^{-2}(t)$, and $v_a$ decrease as $v_a\propto a^{-1}(t)$~\cite[formula 6.8]{gorbrub2}; in addition, the wavelength $\lambda$ increases proportionally to $a(t)$. Therefore the difference between the Hubble constant and $\hu$ at the center of the perturbation $(H-\hu)=\left. dv/dr\right|_{v=0}\propto a^{-2}(t)$. Thus, the isotropic $(H-\hu)$ and anisotropic $h_{ii}$ components of the perturbation in the radiation-dominated stage depend on time in the same way, and their relative magnitude is practically preserved.

Let us now consider the stage of the Universe's expansion dominated by nonrelativistic matter (the so-called dust stage). It began at approximately $z_{eq}\simeq 3600$ (~\cite[formula 4.19]{gorbrub1}, where $\Omega_m=0.306$) and ended at $z\simeq 0.66$,  when the Universe transitioned to accelerated expansion due to a nonzero cosmological constant. In this stage $\hu \propto a^{-3/2}(t)$, while $v_a$ grows: $v_a\propto a^{1/2}(t)$~\cite[formulas 4.22, 7.13]{gorbrub2}, and, of course, as before $\lambda \propto a(t)$. Therefore $(H-\hu)=\left. dv/dr\right|_{v=0}\propto a^{-1/2}(t)$. 

Thus, in the dust stage, in expression~(\ref{27a12}) for the velocity field near the extremum point, the first term (for some fixed $r$) decreases as $a^{-3/2}(t)$, the second as $a^{-1/2}(t)$, and the third as $a^{-2}(t)$. The second term describes the isotropic part of the perturbation and decreases more slowly than all the others. This is precisely why it eventually begins to dominate and, in the case of a density maximum, leads to the replacement of the Hubble expansion of the structure by contraction. The third term of the sum, which describes the anisotropic part of the perturbation, decreases most rapidly. 

We see that the ratio of the anisotropic part of the perturbation to the isotropic part (the third term to the second in~(\ref{27a12})) in the dust stage decreases very rapidly (as $a^{-3/2}(t)$). Since the perturbations evolved from a random Gaussian field, it is reasonable to assume that initially the anisotropic and isotropic components of the perturbation were of the same order of magnitude. We have shown that their equality in the radiation-dominated stage is approximately preserved, but in the dust stage the anisotropic component rapidly dies out. Let us consider for clarity the case when the density extremum is a maximum. Then at some $z_{col}$ the perturbation enters the nonlinear stage and collapses. The ratio of the anisotropic component of the velocity perturbation to its isotropic part by that moment decreases by approximately 
\begin{equation}
\label{27a13}
w=\left(\frac{a_{col}}{a_{eq}}\right)^{3/2}=\left(\frac{z_{eq}+1}{z_{col}+1}\right)^{3/2}
\end{equation}
times. If $z_{col}\simeq 12$, which roughly corresponds to the formation of the first galaxies, then the anisotropic component becomes~$\sim 4600$ times smaller than the isotropic one, i.e., it practically dies out. 

Let us summarize: the only anisotropic (third) term in~(\ref{27a12}) in the dust stage rapidly becomes negligibly small, while the first two describe a spherically symmetric velocity field $\vec v = H(t) \vec r$. We can now formulate the main conclusion of this article: \emph{after the Universe has spent a sufficiently long time in the dust stage, the velocity field of matter near the density extremum point becomes practically isotropic $\vec v = H(t) \vec r$}. 

The physical reasons for such a spherization of velocity are fairly obvious. As we have already done, let us surround the extremum point with a sphere of radius $r=a(t) \tilde R$. If the anisotropic part of the perturbation $h_{ij}$ is absent, the sphere exactly comoves with the matter. If $h_{ij}\ne 0$,  then the matter moves very slowly through the sphere, so that its total flux through it is zero (since $h_{11}+h_{22}+h_{33}=0$). Without loss of generality, let us assume that $h_{11}>0$. Then the matter expands along the OX axis slightly faster than the sphere and passes outward through it. In this case, the matter that ends up on the sphere is that which was previously located closer to the center and therefore had a lower velocity. This leads to an additional decrease of $|h_{11}|$. On the contrary, if $h_{11}<0$, the matter expands along the OX axis slightly more slowly than the sphere, crosses it inward, i.e., arrives from larger radii and therefore has a slightly higher velocity. Since now $h_{11}<0$, this also leads to an additional decrease of $|h_{11}|$.  In essence, this is a special case of the well-known fact that as one approaches the time singularity, the asymmetry of the solution grows~\cite{belinskii1982}: as the Universe expands, it moves away from the singularity — and therefore the asymmetry of the solution decreases.

\section{Discussion}
So far we have considered a perturbation with a spherically symmetric density profile. It is easy to see that the spherization mechanism discussed in this article should also take place in the nonspherical case. The qualitative reasoning given at the end of the previous part demonstrates this most clearly. Indeed, in the nonspherical case as well, if the expansion rate, for example, along the OX axis is higher, this additional expansion carries outward the matter that was initially located closer to the center and had a lower velocity. This effect tends to "equalize" the expansion velocities in different directions and thereby tends to spherize the velocity field. However, the results of the action of the velocity-field spherization effect are completely different in the case of spherical and aspherical density profiles.

In the case of a spherically symmetric density profile of the initial perturbation, not only the particle velocity field $\vec v = H(t) \vec r$, but also the gravitational field turns out to be practically strictly radial, and the distribution of matter density near the extremum is always practically homogeneous ($\rho\simeq{\it const}$). If the density extremum is a minimum, a spherically symmetric void forms. In the case of a density maximum, an interesting possibility arises. As Tolman showed~\cite{tolman1934, bondi1947}, if $\rho={\it const}$, all particles fall toward the center simultaneously, which leads to the formation of a black hole. Of course, in the case of a real distribution, the density at the center of the perturbation will not be exactly constant, and the formation of a black hole will require the fulfillment of certain additional conditions. However, the density of matter during collapse will in any case reach very large values, which the velocity-field spherization effect discussed by us will greatly facilitate.

In the case of a spherically asymmetric density profile, the picture becomes substantially more complicated, since a second effect influencing the sphericity of the perturbation appears: the gravitational field of the perturbation in this case is aspherical. As a result, the acceleration it imparts to matter is also anisotropic. Let us consider for definiteness the case of a density maximum. Generally speaking, the surfaces of constant density near a maximum have the form of ellipsoids. The gravitational field attracts matter toward the center of the ellipsoid; in this case, the greatest acceleration is imparted along its shortest axis \cite{silk1979}. As a result, if the axes of the ellipsoid were not initially equal (i.e., in the case of a spherically asymmetric profile), then the disproportion between the axes only grows: the initially shortest axis becomes even shorter relative to the others, which, in turn, increases the gravitational attraction along this axis. Eventually the ellipsoid collapses into a pancake precisely along its shortest axis.

As we see, even a very small asymmetry of the initial density maximum grows in the final stages of collapse, and ultimately it turns out to be stronger than the weak spherization effect considered in this article: the final stages of collapse are always strongly anisotropic, except in the case of ideal spherical symmetry. However, this same example shows that these two effects have nothing in common. First, in the case of a density maximum their actions are opposite: one of them tends to spherize the velocity field of the perturbation, while the second tends to make it maximally aspherical. Second, the effects depend in completely different ways on the perturbation amplitude. The velocity spherization effect does not depend on the amplitude at all: it operates with constant efficiency from the moment the Universe enters the dust stage $z_{eq}\simeq 3600$. On the contrary, the influence of asphericity becomes especially strong only when the perturbation enters the nonlinear stage, after $z_{col}$, when the asphericity begins to grow rapidly. Before entering the nonlinear stage, the shape of the perturbation remains practically unchanged, and the anisotropic part of the gravitational acceleration turns out to be proportional to the perturbation amplitude \cite{silk1979}. In the dust stage the perturbation amplitude grows proportionally to $a(t)$, and therefore the influence of the gravitational-field asymmetry turns out to be negligibly small at $z\sim z_{eq}$, but us $z_{col}$ is approached it grows rapidly, and at $z>z_{col}$ it completely dominates.

Thus, taking both effects into account, the evolution of the velocity field near the density maximum of an asymmetric perturbation looks as follows. In the radiation-dominated stage, only the amplitude of the density and velocity perturbations changed; the shape and anisotropy of the perturbations remained unchanged. After entering the dust stage $z_{eq}\simeq 3600$ the shape of the perturbation remains practically unchanged, but the velocity field near the density maximum rapidly approaches a homogeneous one owing to the effect considered in this work. If the density perturbation does not possess spherical symmetry, then as it approaches collapse $z_{col}$ (for medium-mass galaxies $z_{col}\simeq 10$) the asymmetry of the gravitational field of the perturbation begins to play an increasingly large role. First it makes the velocity field of the perturbation anisotropic again, and then it leads to a rapid growth of its asymmetry. Ultimately, the perturbation undergoes an asymmetric collapse. 

We arrive at the conclusion that the influence of the effect considered in this work is, generally speaking, comparatively small: for a nonspherical perturbation it reduces to a temporary spherization of the velocity field in the interval $z_{eq}>z>z_{col}$. Spherization can be of significant importance only in the case of the collapse of a perturbation possessing almost ideal spherical symmetry. In this case, near the density maximum not only is the distribution of matter practically homogeneous ($\rho={\it const}$, which is true in the most general case simply by virtue of the properties of an extremum), but also the particle velocity field is practically strictly radial and has the form $\vec v = H(t) \vec r$. As Tolman showed~\cite{tolman1934, bondi1947}, in this case all particles fall toward the center, and simultaneously, which can lead to the formation of a black hole. For this, however, the initial distribution of matter density must be strictly spherically symmetric, which is highly unlikely.

\section{Acknowledgments} The author thanks the University of Helsinki and the Academy of Finland mobility grant 1341541 for financial support of this work.              


\end{document}